\newcommand\BaCu{Ba$_2$CuGe$_2$O$_7$}

\documentclass[twocolumn,prl,aps,showpacs,floatfix,superscriptaddress,preprintnumbers,amsmath,amssymb]{revtex4}

\usepackage{graphicx}

\begin{document}

\title{Double-$k$ phase of the Dzyaloshinskii-Moriya helimagnet Ba$_2$CuGe$_2$O$_7$.}

\author{S. M\"uhlbauer}

\affiliation{Neutron Scattering and Magnetism Group, Institute for Solid State
Physics, ETH Z\"urich, Z\"urich, Switzerland}

\author{S. N. Gvasaliya}

\affiliation{Neutron Scattering and Magnetism Group, Institute for Solid State
Physics, ETH Z\"urich, Z\"urich, Switzerland}

\author{E. Pomjakushina}

\affiliation{Laboratory for Developments and Methods (LDM), Paul Scherrer Institut, Villigen, Switzerland}

\author{A. Zheludev}

\affiliation{Neutron Scattering and Magnetism Group, Institute for Solid State
Physics, ETH Z\"urich, Z\"urich, Switzerland}

\date{\today}

\begin{abstract}

Neutron diffraction is used to re-investigate the magnetic phase diagram of the
non-centrosymmetric tetragonal antiferromagnet Ba$_2$CuGe$_2$O$_7$. A novel
incommensurate double-$k$ magnetic phase is detected near the
commensurate-incommensurate phase transition. This phase is stable only for
field closely aligned with the 4-fold symmetry axis. The results emphasize the
inadequacy of existing theoretical models for this unique material, and points
to additional terms in the Hamiltonian or lattice effects.
\end{abstract}


\maketitle

Dzyaloshinskii-Moriya (DM) antisymmetric exchange interactions
\cite{Dzyaloshinskii:58, Moriya:60} and their role in incommensurate magnetic
structures are once again at the forefront of condensed matter research. One
reason for the renewed interest is the discovery of novel topological skyrmion
phases in the DM helimagnet MnSi and related compounds \cite{Muehlbauer:09b,
Muenzer:09}. The other is the role that DM interactions play in certain
scenarios of the multiferroic effect \cite{Mostovoy:06, Sergienko:06}.
Arguably, no material has attracted more attention simultaneously in both these
contexts than the tetragonal DM helimagnet \BaCu. This compound is one of the
very few systems where DM helimagnetism can coexist with weak-ferromagnetic
behavior \cite{Bogdanov:02}. Moreover, theory suggested that it can support
stable skyrmions, similar to those in MnSi \cite{Bogdanov:02}. \BaCu\ has an
unusual soliton lattice structure in zero field \cite{Zheludev:98PRL}, and
shows a very interesting behavior in applied fields
\cite{Zheludev:97,Zheludev:97b,Zheludev:98PRB}. While the material itself does
not demonstrate spontaneous electric polarization, the isostructural and
magnetically commensurate Ba$_2$CoGe$_2$O$_7$ \cite{Zheludev:03, Sato:03} is,
in fact, an unconventional multiferroic material \cite{Yi:08}.

The most spectacular feature of \BaCu\ is the commensurate-incommensurate (CI)
transition that occurs in this material in applied magnetic fields
\cite{Zheludev:97,Zheludev:98PRB}. Despite the uniqueness and fundamental
importance of this compound, and years of systematic studies, its exact nature
remains unresolved. Theory predicts a continuous transition with a divergent
periodicity, that maps onto the classic Frenkel-Kontorova model
\cite{Frenkel:38}. In obvious contradiction, experiments indicate a
discontinuous reorientation of spins with a poorly defined ``intermediate
state''. The latter has been largely dismissed as a result of defect pinning
\cite{Zheludev:97,Zheludev:98PRB}, phase separation associated with a
spin-flop-like discontinuous transition \cite{Zheludev:99} or a poor alignment
of the applied field relative to the unique 4-fold axis of the tetragonal
structure \cite{Bogdanov:99}. In the present study we show that none of these
explanations apply. In fact, the ``intermediate state'' is a novel double-$k$
incommensurate phase that exists {\it only} for magnetic fields {\it almost
perfectly} aligned along the 4-fold axis. Being very unlike anything predicted
for this material, the double-$k$ structure undermines all existing theoretical
descriptions of \BaCu.

The crystal and magnetic structures of \BaCu\ are reviewed in detail in
Ref.~\onlinecite{Zheludev:98PRB}. The magnetism is due to Cu$^{2+}$ ions that
form a square lattice, cell-centered in the $(a,b)$ plane of the tetragonal
non-centric $P{\overline 4}2_1m$ structure (Fig.~1b in
Ref.~\cite{Zheludev:98PRB}).  Long-range helimagnetic ordering occurs at
$T_\mathrm{N}=3.2$~K. To a good approximation, the low-temperature spin
arrangement is a cycloid with propagation vectors $(1\pm\xi_0, \pm\xi_0,0)$,
$\xi_0=0.027$, and the spins rotating in the $(\pm 1, \mp 1,0)$ planes,
respectively, as illustrated in Fig.~1a of Ref. \cite{Zheludev:98PRB}.
$(1,0,0)$ being the $(\pi,\pi)$ point for the square lattice, nearest-neighbor
spins are aligned almost antiparallel to each other. Due to tetragonal
symmetry, one typically observes two equivalent magnetic domains with
propagation vectors $(1+\xi_0, \xi_0,0)$ and $(1+\xi_0, -\xi_0,0)$,
respectively. Field-cooling in a magnetic field as small as a few tens of Gauss
applied in the $(a,b)$ plane ensures a single-domain structure
\cite{Zheludev:97b}. A more sizable in-plane field tends to re-orient the spin
rotation plane of the entire cycloid to be perpendicular to the field
direction. This is accompanied by a continuous and rather spectacular rotation
of the propagation vector \cite{Zheludev:97b}. The above-mentioned CI
transition occurs in magnetic fields applied along the $c$ axis. It is preceded
by a strong deformation of the planar cycloid, that in applied fields is better
described as a ``solition lattice'' \cite{Zheludev:97,Zheludev:98PRB}. Beyond
$H_c\sim 2.1$~T the system is a commensurate AF with propagation
vector $(1,0,0)$. The focus of the present study is what happens {\it just
before} this commensuration.
\begin{figure}
\includegraphics[width=0.34\textwidth]{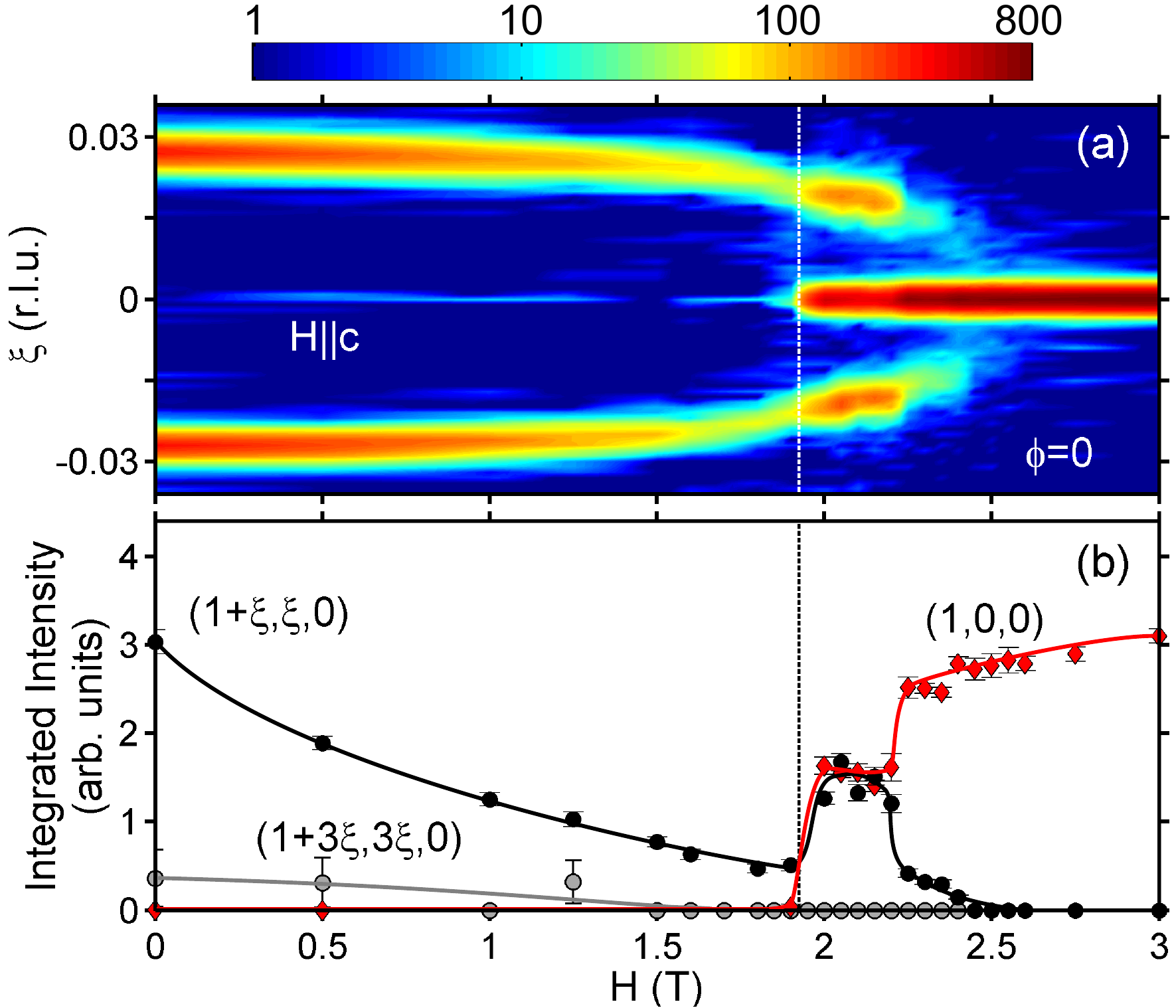}
\caption{(a) Neutron diffraction intensity measured in linear scans along the
$(1+\xi \cos (\phi-\pi/4), \xi \sin (\phi-\pi/4),0)$ reciprocal-space direction
for $\phi=0$ as a function of magnetic field applied along the $c$ axis. (b)
Field dependence of the integrated intensity of the incommensurate satellite
reflections at $\xi$ and $3\xi$ as well as of the commensurate magnetic Bragg
peak at (1,0,0), obtained in Gaussian fits. For the measurements of $3\xi$ the instrumental resolution was relaxed to optimize intensity.} \vspace{-0.04\textwidth}
\label{fig_1}
\end{figure}

A weak point of all previous neutron diffraction studies was a less than
perfect alignment of the magnetic field with respect to the $c$ axis. On the
one hand, this misalignment potentially alters the nature of the phase
transition. On the other hand, measurements and the interpretation of the data
are complicated by a continuous re-distribution of intensity between the two
tetragonal domains. The latter is due to the non-zero in-plane component of the
applied field, and may be history-dependent due to pinning. Thus, the first
step in our strategy to clarify the issue of the ``intermediate state'' was to
perform a new set of diffraction measurements in a sample with an almost
perfect alignment, while diligently following a field-cooling protocol for each
field value.

The data were taken on the TASP 3-axis spectrometer at Paul Scherrer Institut \cite{semadeni:01},
using neutrons of incident wavevector $1.3\,\mathrm{\AA}^{-1}$. The single
crystal sample of mass 1.65~g was grown using the floating-zone technique and
mounted in a split-coil cryomagnet. A misalignment of less that $0.5^\circ$ between the nominally vertical
field and the crystallographic $c$ axis at low temperatures was achieved. The data were
taken after field-cooling from 6~K to 1.6~K, in sets of
elastic scans along the $(1+\xi \cos (\phi-\pi/4), \xi \sin (\phi-\pi/4),0)$
directions in reciprocal space, for $\phi=0$ and $\phi=\pi/2$. Here we have
adopted the notation used in Ref.~\cite{Zheludev:97b}, where $\phi$ is the
angle between the $(1,1,0)$ direction and the propagation vector. The
background, primarily due to double scattering \cite{Zheludev:97}, was subtracted from all low-temperature
scans.

\begin{figure}
\includegraphics[width=0.35\textwidth]{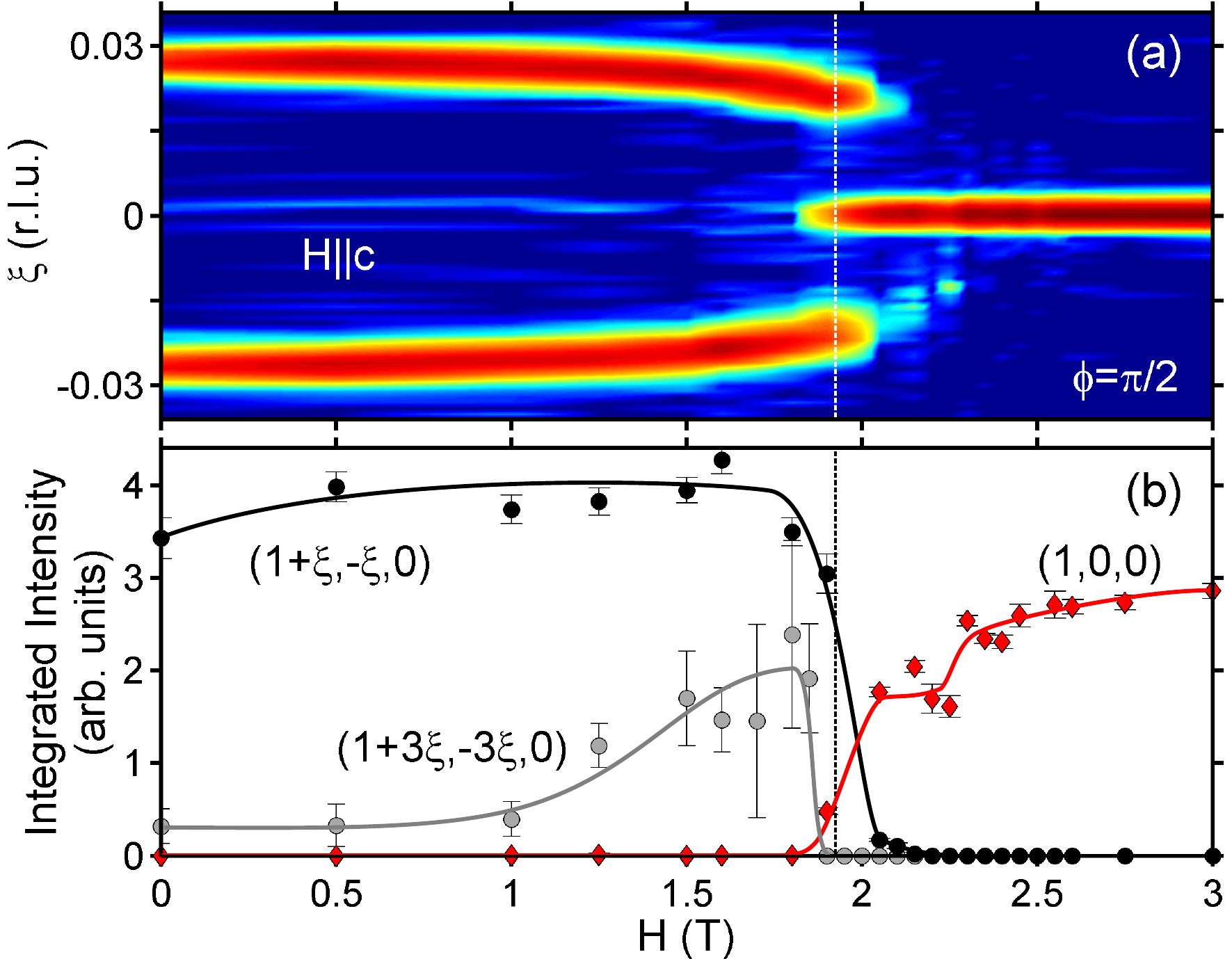}
\caption{Neutron diffraction data as a function of magnetic field applied along
the $c$ axis for $\phi=\pi/2$, analogous to Fig.~\ref{fig_1}. The intensity
scale as given in Fig.~\ref{fig_1} is used.} \label{fig_2}
\vspace{-0.03\textwidth}\end{figure}

The result of these measurements are summarized in Figs.~\ref{fig_1} and ~\ref{fig_2}. With increasing
magnetic field, we observe a gradual change of the magnetic
period $\zeta$ and a shift of intensity from the $\phi=0$ domain to that with
$\phi=\pi/2$.  Even at $H=1.7$~T a single-domain state is not achieved,  though
most of the intensity has shifted to $\phi=\pi/2$. This confirms an
almost-perfect alignment of our sample relative to the $c$ axis. A dramatic
redistribution of intensity is seen at $H_1=1.9$~T. At this point the
incommensurate peaks in the $\phi=\pi/2$ scans disappear. At the same time,
those in $\phi=0$ scans abruptly gain intensity. In other words, at $H_1$ the
preferred incommensurate propagation vector abruptly rotates by $\pi/2$.
Moreover, an additional peak appears in all scans in the commensurate $(1,0,0)$
position. The incommensurate peaks persist until it disappears at the CI transition at
$H_c\sim2.4$~T. We identify the
field range $H_1<H<H_c$ with the previously reported intermediate state.

The abrupt rotation of the incommensurate vector at $H_1$ suggests that this
state is {\it not} a phase-separated mixture of the commensurate and the
low-field cycloidal phases, as previously speculated, but a distinct double-$k$
incommensurate structure. The same conclusion can be drawn from the behavior of
higher-order harmonics, such as the ones at $(1+3\xi \cos (\phi-\pi/4), 3\xi
\sin (\phi-\pi/4),0)$. At $H<H_1$, the 3rd harmonic increases with field
(Fig.~\ref{fig_2} b), due to a distortion of the planar cycloid and the
development of a soliton lattice \cite{Zheludev:97}. In contrast, for
$H_1<H<H_c$ the harmonics are absent. The incommensurate modulation is purely
sinusoidal, without distortion, and thereby distinct from that below $H_1$.

A distinct incommensurate ``cone'' phase preceding to the CI transition was
indeed predicted theoretically, for fields applied at an angle $\alpha>0$ with
respect to the 4-fold axis \cite{Bogdanov:99}. This phase is supposedly absent
for $\alpha=0$, and its field range progressively increases with increasing
$\alpha$. To verify that the observed double-$k$ structure is not due to a
residual $\alpha<0.5^\circ$ field misalignment, we systematically investigated
the behavior in canted fields.

\begin{figure}
\includegraphics[width=0.35\textwidth]{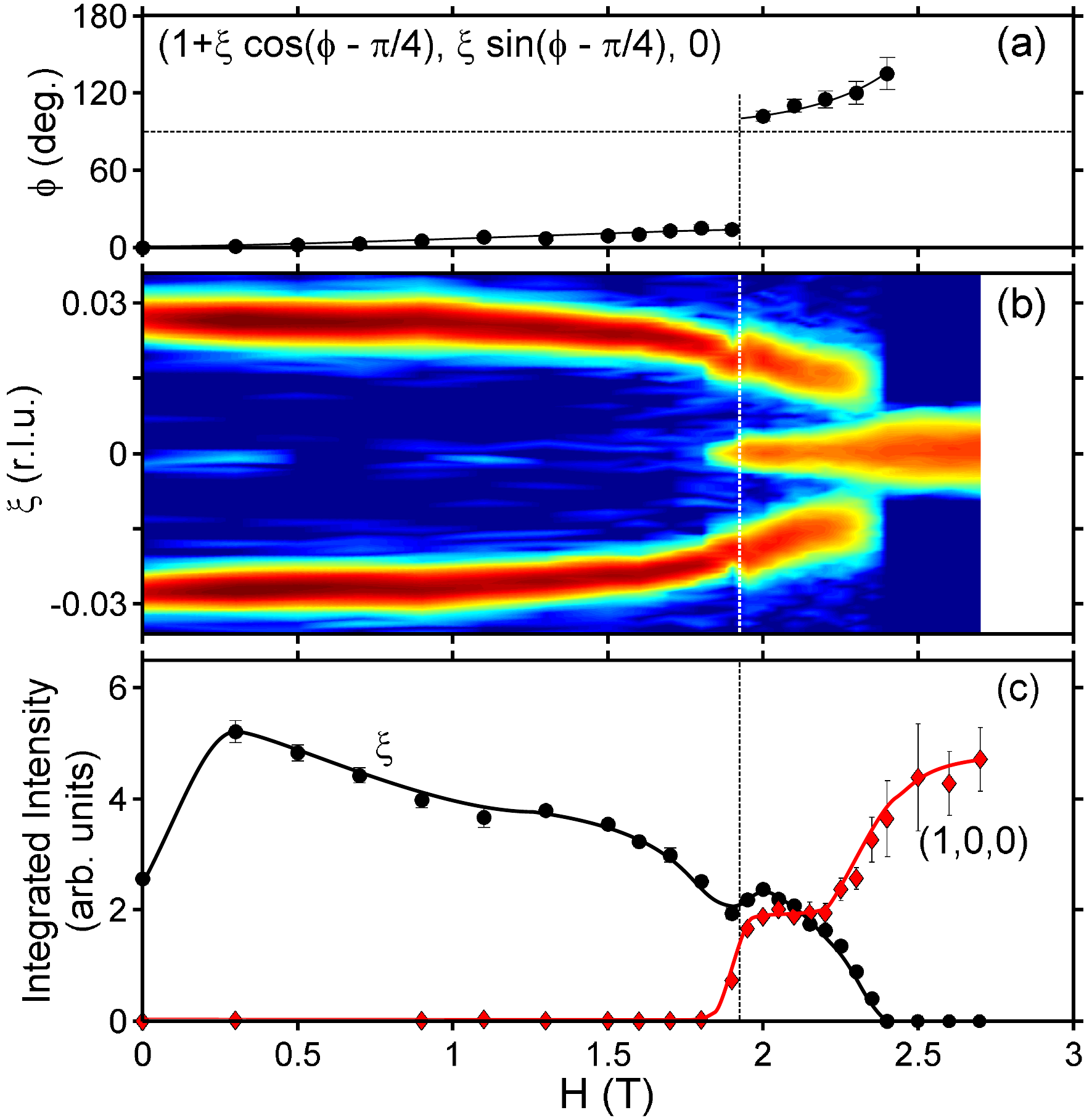}
\caption{Neutron diffraction data as a function of magnetic field
for $\alpha=5^{\circ}$. (a) Magnetic field dependence of $\phi$. (b)
Neutron diffraction intensity measured in linear scans along the
$(1+\xi \cos (\phi-\pi/4), \xi \sin (\phi-\pi/4),0)$
reciprocal-space direction. The intensity scale as given in
Fig.~\ref{fig_1} is used. (c) Field dependence of the integrated
intensity of the incommensurate satellite reflections at $\xi$ as
well as of the commensurate magnetic Bragg peak at (1,0,0), as
obtained in gaussian fits.} \label{fig_3}
\vspace{-0.03\textwidth}\end{figure}

In the first such measurement the field was applied at an angle
$\alpha=5^\circ$ relative to the $c$ axis, in the $(1,0,0)$ plane.
Despite the much larger misalignment angle, we observe an almost
unchanged behavior. These data are summarized in
Fig.~\ref{fig_3} a--c. Due to a sizable in-plane field component,
only one domain, with $\phi\sim0$, survives beyond 0.5~T applied
field. The same in-plane component induces a gradual rotation of the
incommensurate peak around the $(1,0,0)$ position, $\phi$ changing
from $\phi=0$ at $H=0$ to $\phi\sim 15^\circ$ at $H=1.9$~T in
agreement with Ref.~\cite{Zheludev:97b}. The direction of the
propagation vector was determined by centering the incommensurate
peaks at each field. The $\xi$-scans that make up the false-color
plot in Fig.~\ref{fig_3} b are thus performed along the $(1+\xi \cos
(\phi-\pi/4), \xi \sin (\phi-\pi/4),0)$ direction with a
field-dependent $\phi$. As for zero tilt, at $H=H_1\sim 1.9$~T, the
propagation vector abruptly rotates by precisely 90 degrees
($\phi\rightarrow \phi+\pi/2$ ), and a commensurate component
appears. The incommensurate peak vanishes at $H_c\sim 2.4$~T.

\begin{figure}
\includegraphics[width=0.35\textwidth]{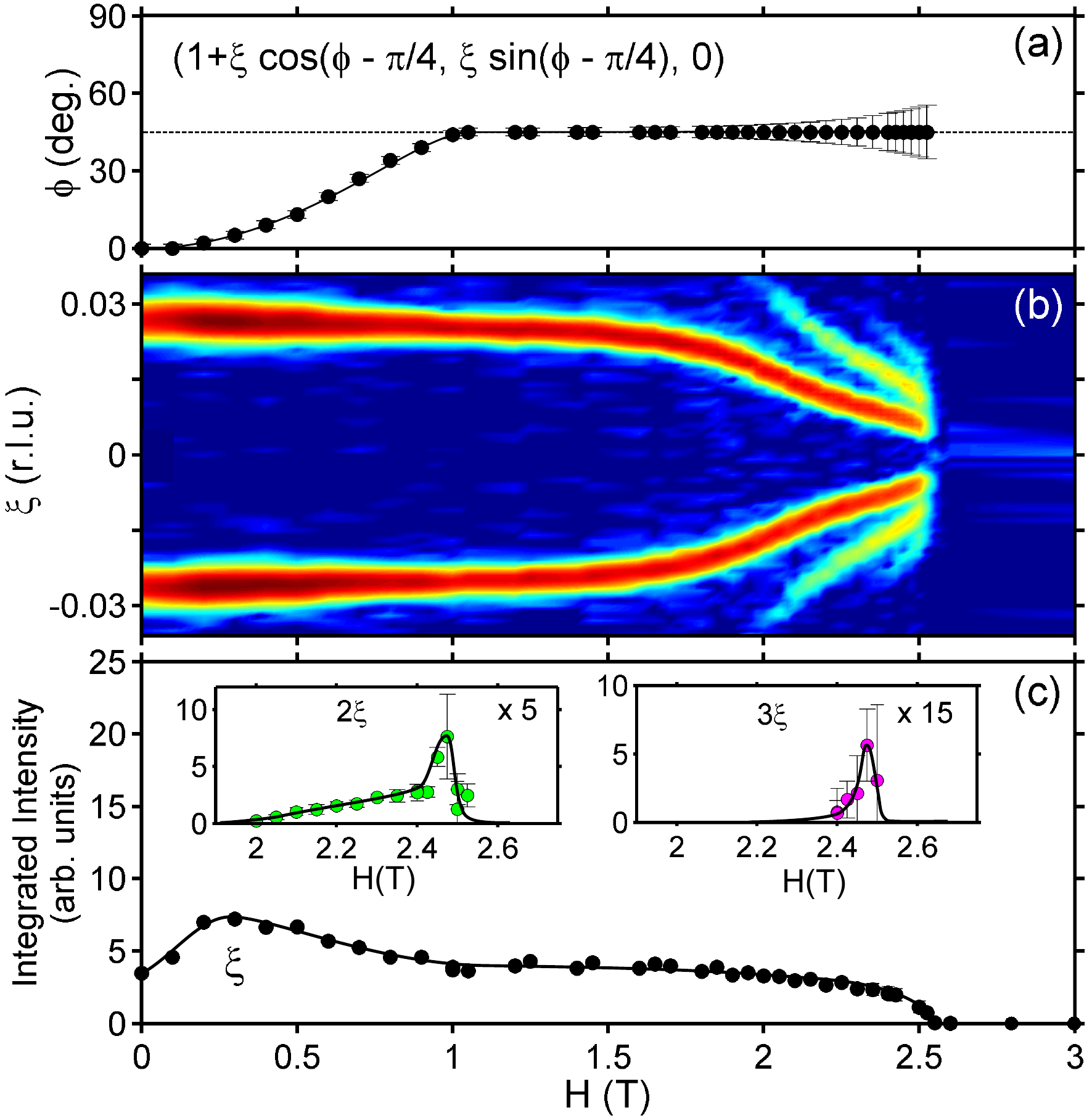}
\caption{Neutron diffraction data as a function of magnetic field
for $\alpha=15^{\circ}$, analogous to Fig.~\ref{fig_3}. The
intensity scale as given in Fig.~\ref{fig_1} is used.} \label{fig_4}
\vspace{-0.03\textwidth}\end{figure}

A key finding of this study is the double-$k$ phase is not enhanced by a field
misalignment with respect to the $c$ axis, but is actually absent at large
$\alpha$ angles. Data collected for $\alpha=15^\circ$ are shown in
Fig.~\ref{fig_4}. Due to a much larger in-plane field component, the
propagation vector rotates more rapidly and completely aligns itself with the
crystallographic $(1,0,0)$ direction: $\phi \rightarrow \pi/4$ for $H\gtrsim
1$~T. This gradual re-orientation is fully consistent with previous studies
\cite{Zheludev:97b}. No abrupt re-orientation of the propagation vector is
observed at any field. The behavior is continuous all the way to the CI
transition at $H_c\sim 2.5$~T, with no sign of an additional phase at lower
fields.

In contrast to small-$\alpha$ geometries, where the CI transition occurs in a
sinusoidally modulated phase with no higher-order harmonics, for
$\alpha=15^\circ$ we observe a strong distortion of sinusoidal modulation.
Indeed, as clearly seen in Fig.~\ref{fig_4}, the third ($3\xi$), as well as
the previously unobserved 2nd ($2\xi$) harmonics, drastically increase in
intensity as the CI transition is approached.

Note that in Fig.~\ref{fig_4}b, at $H> H_c$, the commensurate peak appears to
be absent. This, however, is an instrumental effect. It stands to reason that
in the commensurate AF phase, spins are aligned perpendicular to
the field, i.e., along the $(1,0,0)$ direction. Neutrons are only scattered by
spin components that are transverse to the scattering vector. As a result, the
$(1,0,0)$ AF Bragg peak is extinguished by this polarization effect. Due to the
restrictive geometry of a split-coil cryomagnet, measurements at equivalent
wave vectors with more favorable polarization were not feasible.

Preliminary data (not show) indicate virtually identical behaviors for fields
confined to the $(1,1,0)$ plane \cite{Muehlbauer:unpublished}. For $\alpha=5^\circ$ the propagation vector jumps by $90^\circ$ at
$H_1\sim1.9$~T, while for $\alpha=15^\circ$ no reorientation is observed and
the peak remain at $\phi=0$ all the way to the CI transition. We can therefore
summarize all diffraction results as follows. In field applied almost parallel
to the 4-fold axis, a novel sinusoidally modulated double-$k$ phase appears
just before the CI transition. When the field deviates from the high-symmetry
direction, the double-$k$ phase is absent. In this case the CI transition
occurs in a complexly distorted single-$k$ phase that continuously evolves from
the zero field structure.

\begin{figure}
\includegraphics[width=0.48\textwidth]{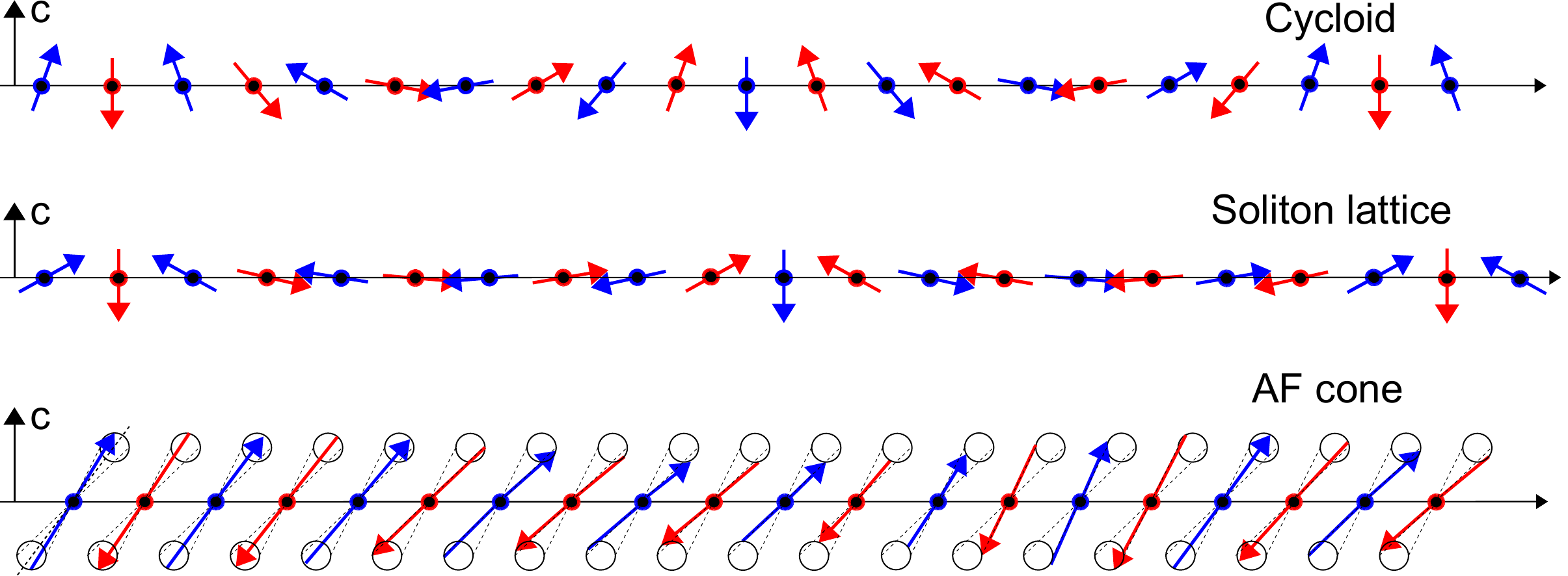}
\caption{Proposed magnetic structure of \BaCu for the almost AF
cycloid, the soliton lattice and the AF-cone phase.} \label{fig_5}
\vspace{-0.03\textwidth}\end{figure}

Even the limited diffraction data allow us to make a good guess regarding the
nature of the novel double-$k$ phase. Since it has a commensurate {\it
antiferromagnetic} component, the double-$k$ structure is clearly {\it not} the
conical phase discussed in Ref.~\cite{Bogdanov:99}. Instead, we propose that it
can be approximated as an {\it antiferromagnetic cone} structure depicted in
Fig.~\ref{fig_5}. The spins are mostly aligned in a commensurate AF pattern in
the $(a,b)$ plane. In addition, there is a small incommensurate precession of
transverse spin components. We will assume that, due to tetragonal symmetry,
this structure can freely rotate in the $(a,b)$ plane, just like the soliton
lattice \cite{Zheludev:97b}. Similarly, we can expect that the orientation of
the spin rotation plane is uniquely coupled to the propagation vector.

In a field, any AF structure will always favor  a spin-flop configuration. In
particular, the dominant commensurate AF spin component of the proposed AF-cone
structure will tend to align itself perpendicular to any in-plane component of
the applied field. As a result, the spin precession plane will align {\it
parallel} to the in-plane field component. The expected behavior for the planar
soliton-lattice phase (Fig.~\ref{fig_5}b) is quite different. A small in-plane
field will tend to align the spin rotation plane {\it perpendicular} to itself
\cite{Zheludev:97b}.  Thus the observed abrupt $90^\circ$ rotation of the
propagation vector at $H_1$ may signify a transition from the planar
soliton-lattice phase at low fields to a AF-conical state at high fields.

The proposed model for the double-$k$ structure also gives a handwaving
explanation of the observed absence of higher-order harmonics. Indeed, for the
planar state realized below $H_1$, a distortion and formation of a soliton
lattice is the only way it can respond to a field applied in the plane of spin
rotation. In contrast, the AF cone state, can take advantage of the Zeeman
energy by a canting of its dominant AF-commensurate component, leaving the
small incommensurate modulation unperturbed.

A further understanding of the observed double-$k$ phase will require new
theoretical input. Existing models fail to predict it, and are clearly missing
an important component of the physics of \BaCu. The new phase appears close to
the CI transition, where the soliton lattice of the original model becomes
infinitely soft. Therefore, the missing terms in the Hamiltonian need not be
particularly strong to qualitatively alter the energy landscape. We can suggest
two potential candidates. First, many of previous theories focused only on the
component of the DM vector $\mathbf{D}$ that lies in the tetragonal $(a,b)$
plane. Due to crystallographic symmetry, this component is uniform from one
Cu-Cu bond to the next, and is responsible for all the spiral structures in
\BaCu. The component of $\mathbf{D}$ that is parallel to the four-fold axis
(Fig.~1b in Ref.~\cite{Zheludev:98PRB}) has often been ignored. This component
is sign-alternating and favors a commensurate weak-ferromagnetic canting of all
spins in one direction \cite{Bogdanov:04}. This term is known to be dominant in
the similar K$_2$V$_3$O$_8$ system \cite{Lumsden2001}. If active in
\BaCu, it will distort any planar soliton-lattice phase, potentially giving
rise to {\it even}-order harmonics, as those observed in this work for large
$\alpha$. Note that distortions due to magnetic field itself produce only odd
harmonics \cite{Zheludev:97}. We can speculate that this term could also play a
role in stabilizing the double-$k$ structure.

The other potential cause of the new phase is lattice effect. Indeed, all
previous models were based on a continuous mapping of the original Heisenberg
spin Hamiltonian. Near the CI transition though, such a description is bound to
break down. In this theory, even as the distance between solitons increases
near $H_c$, the {\it width} of each soliton decreases. Near the critical field
it will become comparable to the lattice spacing, at which point the theory
will lose self-consistency.

In summary, our results show that there is more to the physics of \BaCu\ than
previously thought. This circumstance may have a direct bearing on the
stability of the predicted (but never observed) skyrmion lattice in this
material \cite{Bogdanov:02,Bogdanov:04}, as well as the multiferroic effect.

This work is partially supported by the Swiss National Fund  through Project 6
of MANEP.  We thank  C. Pfleiderer, K. Conder, J. Gavilano, M. Laver, U. R\"ossler and A. Bogdanov for support
and stimulating discussions. Technical support from M. Zolliker and M.
Bartkowiak is greatfully acknowledged.

\end{document}